\documentclass[12pt]{article}
\usepackage{epsfig}

\newcommand{\be}{\begin{eqnarray}}
\newcommand{\ee}{\end{eqnarray}}

\begin{document}

\bibliographystyle{unsrt}
\footskip 1.0cm

\thispagestyle{empty}
\begin{flushright}
INT--PUB 04--03
\end{flushright}
\vspace{0.3in}

\begin{center}{\Large \bf {Forward Rapidity Hadron Production 
in Deuteron Gold Collisions from Valence Quarks}}\\

\vspace{1.8in}
{\large  Jamal Jalilian-Marian}\\

\vspace{.2in}
{\it Institute for Nuclear Theory, University of Washington, 
Seattle, WA 98195\\}

\end{center}

\vspace*{25mm}

\begin{abstract}

\noindent We consider hadron production in deuteron gold collisions at 
RHIC in the forward rapidity region. Treating the target nucleus as 
a Color Glass Condensate and the projectile deuteron as a dilute
system of valence quarks, we obtain good agreement with the BRAHMS
minimum bias data on charged hadron production in the forward rapidity 
($y=3.2$) and low $p_t$ region. We provide predictions for neutral 
pion production in minimum bias deuteron gold collisions in the 
forward rapidity region, $y=3.8$, measured by the STAR collaboration 
at RHIC.

\end{abstract}
\newpage

\section{Introduction}

The recent observation of the suppression of the charged hadron spectra
in the forward rapidity region \cite{brahms} at RHIC, and its centrality 
dependence has generated a lot of interest and excitement in the high 
energy heavy ion community. The suppression of particle spectra and 
disappearance of the Cronin effect at forward rapidity had been 
predicted \cite{kkt,aaksw} in the Color Glass Condensate formalism 
\cite{cgc}, unlike the 
more conventional models which predicted a stronger enhancement of
the spectra in the forward rapidity as compared with mid rapidity \cite{iv}.
The forward rapidity data strongly suggest that the high gluon density 
region of QCD phase space, the Color Glass Condensate, has been 
observed at RHIC. While the Color Glass Condensate has been successful 
in predicting some global features of the data, such
as multiplicities and their energy, rapidity and centrality dependence
in gold gold and deuteron gold collisions \cite{kln} as well as 
qualitative predictions \cite{kkt,aaksw} of the suppression of the 
hadron spectra and their centrality dependence, there has not been a 
quantitative analysis of the forward rapidity hadron spectra using the
Color Glass Condensate formalism. Here, for the first time, we provide a 
limited, but quantitative analysis of the low $p_t$, forward rapidity 
RHIC data, using the Color Glass Condensate formalism.

As emphasized in \cite{adjjm}, the forward rapidity region at RHIC
is the best kinematic region to look for the signatures of the Color
Glass Condensate since this is the region where one probes the smallest
$x$ in the target nucleus so that the Color Glass Condensate will be
manifest more strongly in this kinematic region. Also, forward rapidity 
deuteron gold collisions are ideal since one does not have the final 
state (Quark Gluon Plasma) interactions, present in mid rapidity heavy 
ion collisions. 

It is important to note that at very forward rapidities ($y > 3$), one
is probing the small $x$ region of the nucleus and the {\it large} $x$
region of the deuteron projectile. For example, at $y \sim 3.2$ and 
$p_t \sim 2$, the deuteron wave function at $x \sim 0.25$ is probed. 
This is the region where valence quarks dominate over gluons and sea 
quarks in the deuteron wave function while the relevant $x$ for the 
target nucleus is $x \sim 4\times 10^{-4}$ so that gluons are the 
dominant parton species in the target nucleus. 

In this work, we concentrate in the very forward
rapidity region ($y \ge 3.2$) so that the target nucleus is treated as
a Color Glass Condensate while only the valence quarks in the projectile 
deuteron are included. We use the results of \cite{adjjm} for the scattering
of valence quarks on a Color Glass Condensate to calculate charged hadron
and neutral pion $p_t$ spectra as well as the nuclear modification factor
$R_{dA}$ in the forward rapidity region at RHIC.

\section{Scattering of quarks on a Color Glass Condensate}

In \cite{adjjm}, scattering of quark on a target described as a Color
Glass Condensate was considered and the scattering cross section was
calculated. The incoming quark is taken to be massless and carries zero
transverse momentum. The scattering cross section is given by

\begin{eqnarray}
q^- {d\sigma^{qA\rightarrow qX} \over d^2q_t dq^-} =- {1 \over(2\pi)^2}\,
q^- \, \delta(q^--p^-) \int d^2b_t \, d^2 r_t e^{i q_t\cdot r_t}
\sigma_{dipole}(r_t,b_t,x_g)
\label{eq:qAcs}
\end{eqnarray}

Here, $p^-$ is the light cone energy of the incoming quark while $q^-$
is the light cone energy of the outgoing quark with transverse momentum
$q_t$ and  $\sigma_{dipole}$ is the cross section for scattering of a 
quark anti-quark dipole on a target described as a Color Glass Condensate.
The dipole cross section satisfies the non-linear JIMWLK \cite{jimwlk}
(BK \cite{bk} at large $N_c$)
evolution equation. To relate this quark-nucleus (proton) target 
scattering cross section to hadron production in deuteron gold 
collisions, we convolute this
cross section with the quark distribution function in a deuteron and 
quark-hadron fragmentation function. We get  
\begin{eqnarray}
&&{d\sigma^{dA\rightarrow h(y,k_t)X} \over dyd^2k_t}= - 
{1\over (2\pi)^2}\sqrt{k_t^2 \over s}\,e^y\,\int_{z_{min}}^1 dz\,
q_d\left(x_q\right)\,D_{q/h}(z) \nonumber\\
&&\qquad\times 
\int d^2 b_t \, d^2 r_t e^{i k_t \cdot r_t/z}
\sigma_{dipole}(r_t,b_t,x_g)
\label{eq:csrap}
\end{eqnarray}
where we have used the following kinematical relations $x_q=k_\perp
e^y/z\sqrt{s}$, $x_g=k_\perp e^{-y}/z\sqrt{s}$, $z_{\rm min}= k_\perp
e^y/\sqrt{s}$ ($\sqrt{s}$ is the center of mass energy) while $y$ and
$k_t$ are the rapidity and transverse momentum of the measured hadron.
Both the quark distribution and fragmentation functions depend on a 
factorization scale $Q_f^2$ which is not written out explicitly. In our
calculation, we set $Q_f = k_t$ where $k_t$ is the transverse momentum
of the observed particle. Eq. (\ref{eq:csrap}) is the formula used in 
this work to calculate the hadron spectra at forward rapidities in 
deuteron gold collisions.

We use the LO GRV98 quark distribution function \cite{grv98} and the 
LO KKP quark hadron fragmentation function \cite{kkp}. It should be noted 
that there is no available fragmentation function for negatively charged 
hadrons so therefore we use
the fragmentation function for $(h^+ + h^ -)/2$. However, in the
transverse momentum range considered here ($p_t < 2 GeV$), this should 
not make a sizable difference. Since we are sensitive only to the 
large $x_q$ ($ \sim 0.2-0.5$) region of the deuteron wave function, there
is practically no sensitivity to the choice of the quark distribution 
function. Also, the nuclear modification of valence quarks 
in the deuteron wave function is minimal in this $x_q$ range and is 
therefore neglected here.

To proceed further, we need to know the dipole cross section 
$\sigma_{dipole}$ which satisfies the JIMWLK equation. This is a 
very complicated functional equation \cite{hw} which simplifies 
in the large $N_c$ limit, known as the BK equation. The BK equation 
for the dipole cross section has been numerically solved by various 
people in different limits \cite{solbk}. Iancu, Itakura and
Munier proposed a parameterization of the dipole cross section which 
has all the properties of the solution to the BK equation and used it 
to fit the HERA data on the proton structure function $F_2$ \cite{iim}. 
This is 
a very simple and economical parameterization (it basically has three 
free parameters, in addition to 
the light quark mass) which does an excellent job of describing the HERA 
data at $x < 0.01$. Therefore, we use this parameterization in this work. 
The dipole cross section is given by 

\be
\int d^2 b_t \, \sigma_{dipole}(x_g,b_t,r_t) \equiv 
2 \pi R^2  \,\,{\cal N} \,(x_g, r_t Q_s)
\label{eq:cs_iim}
\ee
where 
\be
{\cal N} \,(x_g, r_t Q_s) = 1 - e^{-a \ln^2 b\,r_t Q_s}  \,\,\,\,\,\,\,\,\,\,
r_t Q_s > 2 
\nonumber
\ee
and 
\be
{\cal N} \,(x_g, r_t Q_s) = {\cal N}_0 \exp \Bigg\{ 2 \ln ({r_t Q_s \over 2})
\bigg [\gamma_s + {\ln 2/r_t Q_s \over \kappa \lambda \ln 1/x_g} \bigg] \Bigg\}
\,\,\,\,\,\,\,\,\,\,
r_t Q_s < 2 
\label{eq:cs_param}
\ee
The constants $a, b$ are determined by matching the solutions at $r_t Q_s =2$
and $\gamma_s = 0.63$ and $\kappa = 9.9$ are determined from LO BFKL.
The form of the proton saturation scale $Q_s^2$ is taken to be 
$Q_s^2 \equiv (x_0/x)^{\lambda} \, GeV^2$ with $x_0, \lambda, {\cal N}_0$
determined from fitting the HERA data on proton structure function $F_2$.
We refer the reader to \cite{iim} for details of the fit. In case of a
nucleus, we have $R_{A} = 1.1 \, A^{1/3} R_{p}$ and 
use the value of the minimum bias saturation scale given
in \cite{kln} ($Q^2_{s, {\it min \, bias}} = 0.95 \,GeV^2$ at $x = 0.01$), 
with the same $x$ dependence as in a proton given above.

\begin{figure}[htp]
\centering
\setlength{\epsfxsize=16cm}
\centerline{\epsffile{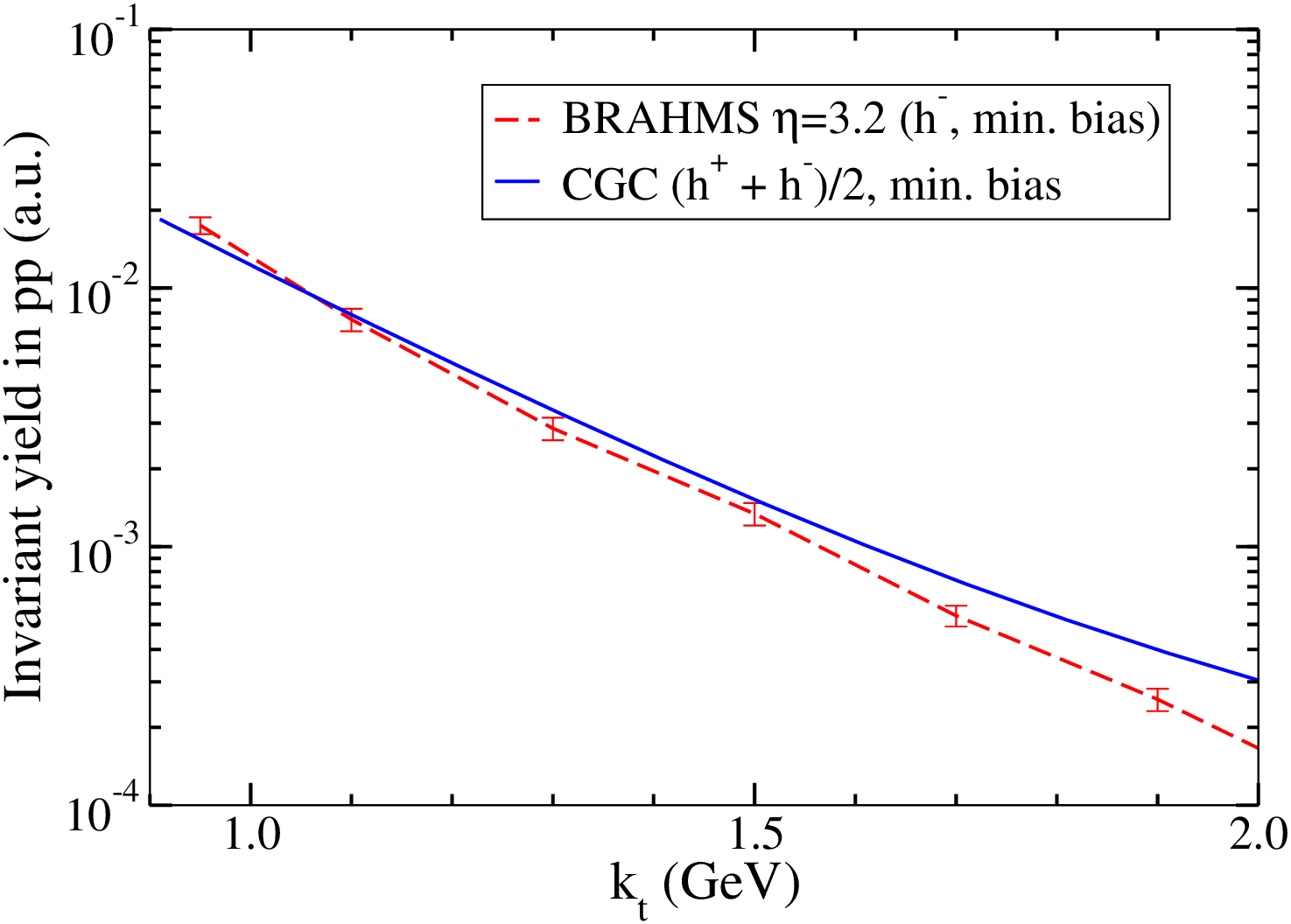}}
\caption{The invariant yield of charged hadrons at $y=3.2$ at RHIC in 
proton-proton collisions at $\sqrt{s}=200$ GeV. The normalization 
is a fit to the data.}
\label{fig:cs_pp_32}
\end{figure}

\begin{figure}[htp]
\centering
\setlength{\epsfxsize=16cm}
\centerline{\epsffile{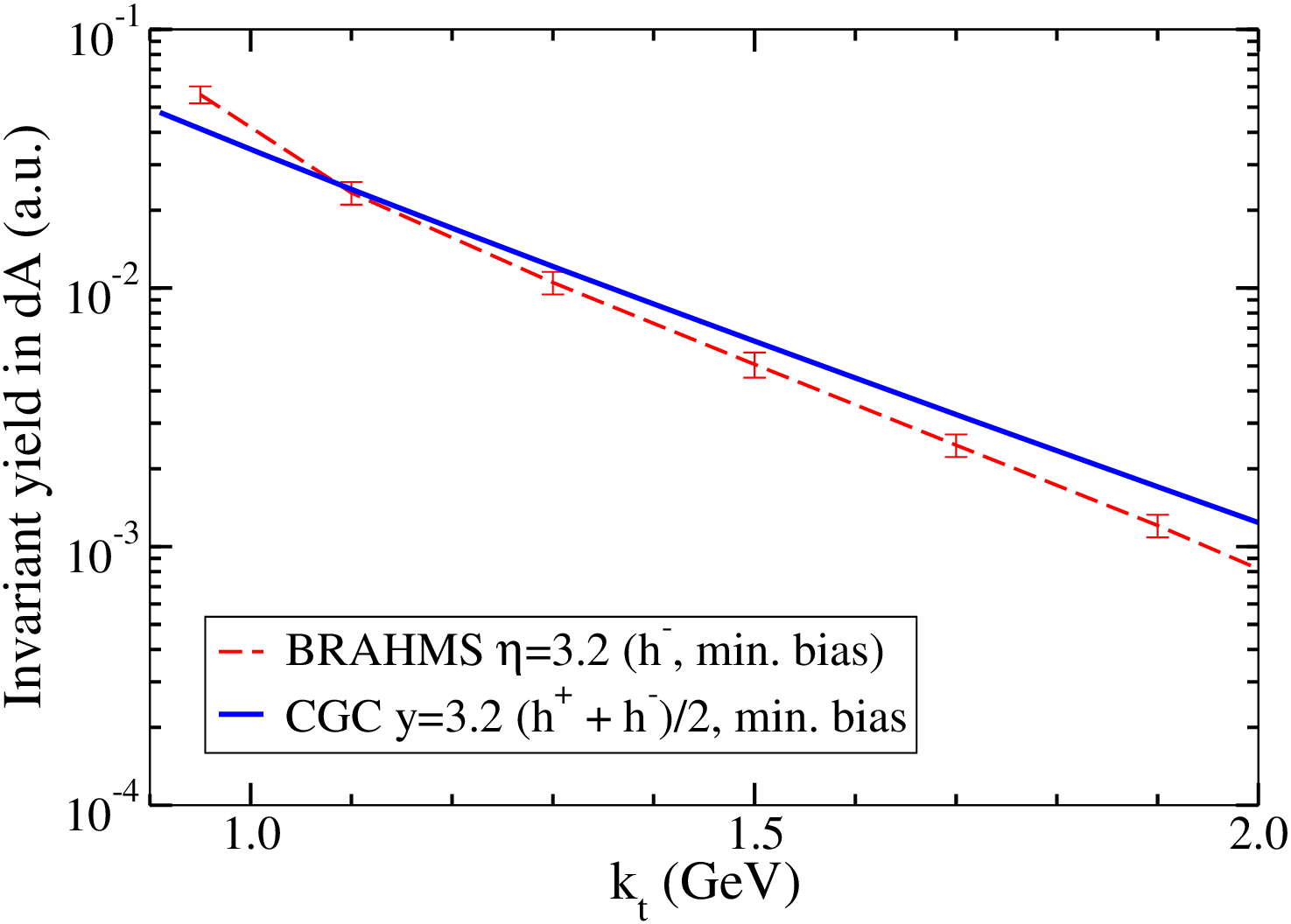}}
\caption{The invariant yield of charged hadrons at $y=3.2$ at RHIC in 
minimum bias deuteron-nucleus collisions at $\sqrt{s}=200$ GeV.}
\label{fig:cs_dA_32}
\end{figure}

In Fig. (\ref{fig:cs_pp_32}), we show our results for charged hadron 
production at $y=3.2$ in proton proton collisions in
arbitrary units while in Fig. (\ref{fig:cs_dA_32}), our results
for charged hadron production in deuteron gold collisions is shown. 
Both figures are for minimum bias events. The proton-proton cross section
is normalized to the data at $k_t = 1.1$ GeV by multiplying by a $K$
factor of $2.57$. To get the normalization of the $dA$ spectra, we again 
multiply by a $K$ factor which is slightly less than the $K=1.8$ factor 
for the proton-proton case. This is not unreasonable since Next to Leading 
Order corrections are typically large at low $k_t$ and that the $K$ factors
used can be different for nuclei \cite{ina}. However, the ratio $R_dA$ is 
calculated {\it without} any $K$ factor. The agreement with 
the slope of the spectra for $k_t < 2$ GeV is quite reasonable specially 
since there are {\it no free parameters} in this calculation. 

\begin{figure}[hbp]
\centering
\setlength{\epsfxsize=16cm}
\centerline{\epsffile{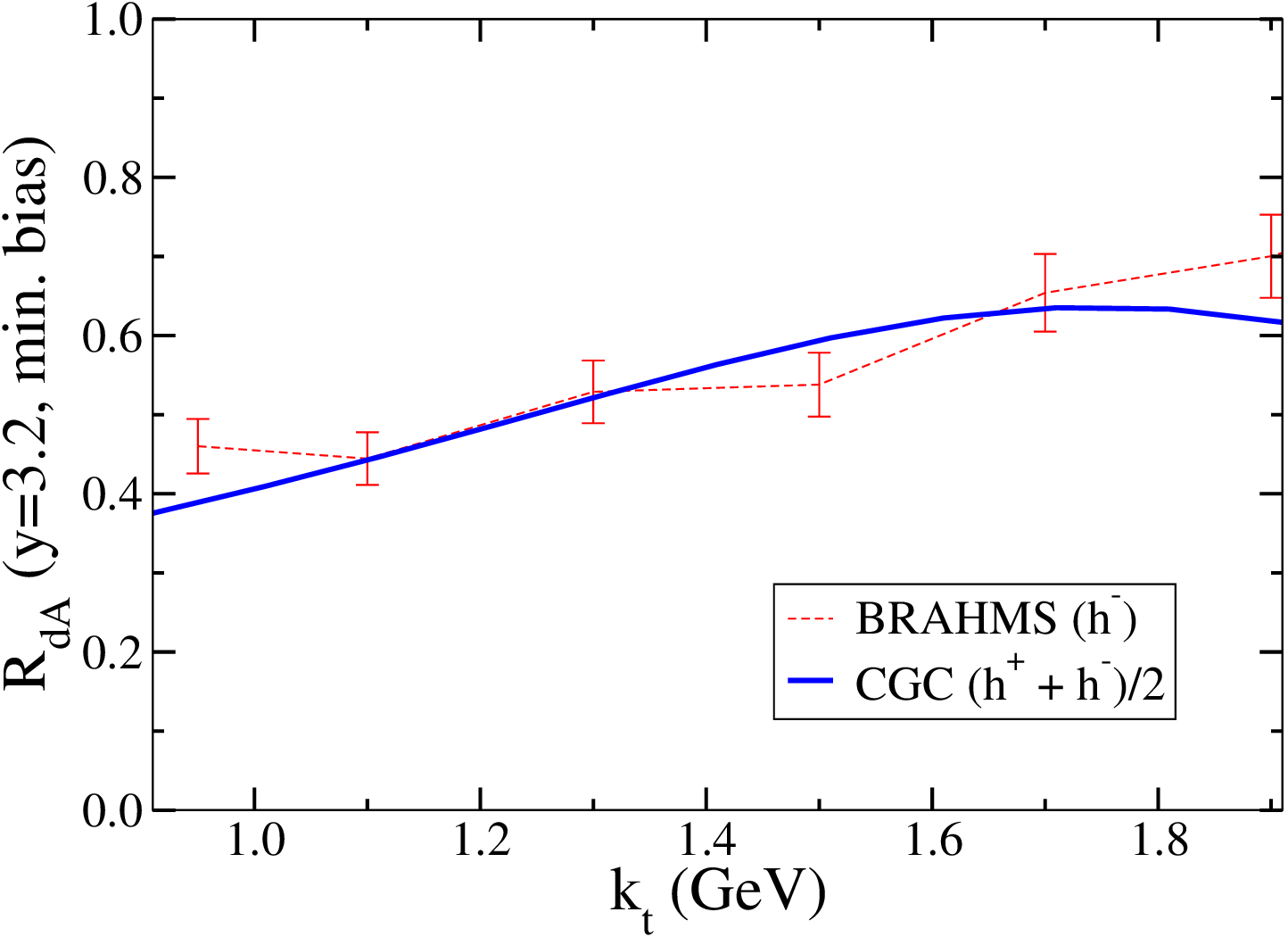}}
\caption{The nuclear modification factor for charged hadrons at $y=3.2$ 
in minimum bias deuteron-nucleus collisions at RHIC, $\sqrt{s}=200$ GeV.}
\label{fig:R_dA_32}
\end{figure}

In Fig. (\ref{fig:R_dA_32}), we show the nuclear modification factor $R_{dA}$ 
for hadron production in deuteron gold collisions at $y=3.2$. We define
\be
R_{dA}\equiv {{d\sigma^{dA \rightarrow h X} \over dy d^2k_t} \over
2\,A\, {d\sigma^{pp \rightarrow h X} \over dy d^2k_t}}
\label{eq:R_dA}
\ee
Again, the agreement with the data at low $k_t$ is quite good but the higher 
$k_t$ points start to show a deviation. This is discussed in more detail 
later. We emphasis the point that the ratio $R_dA$ is calculated without
multiplying by any $K$ factor and with no free parameters.

It should be noted that the BRAHMS data is for negatively charged
hadrons while our calculations are done for the average of positively
and negatively charged hadrons since fragmentation functions for
negatively charged hadrons are not available. Recently, 
Guzey et al. \cite{guzey} investigated the dependence of the suppression
and showed that isospin symmetry considerations can make a huge effect
on the observed suppression. This effect is most pronounced in the 
$k_t > 2$ GeV and can affect our results by $(15-20) \%$ in the kinematic
region we cover.

\begin{figure}[hbp]
\centering
\setlength{\epsfxsize=16cm}
\centerline{\epsffile{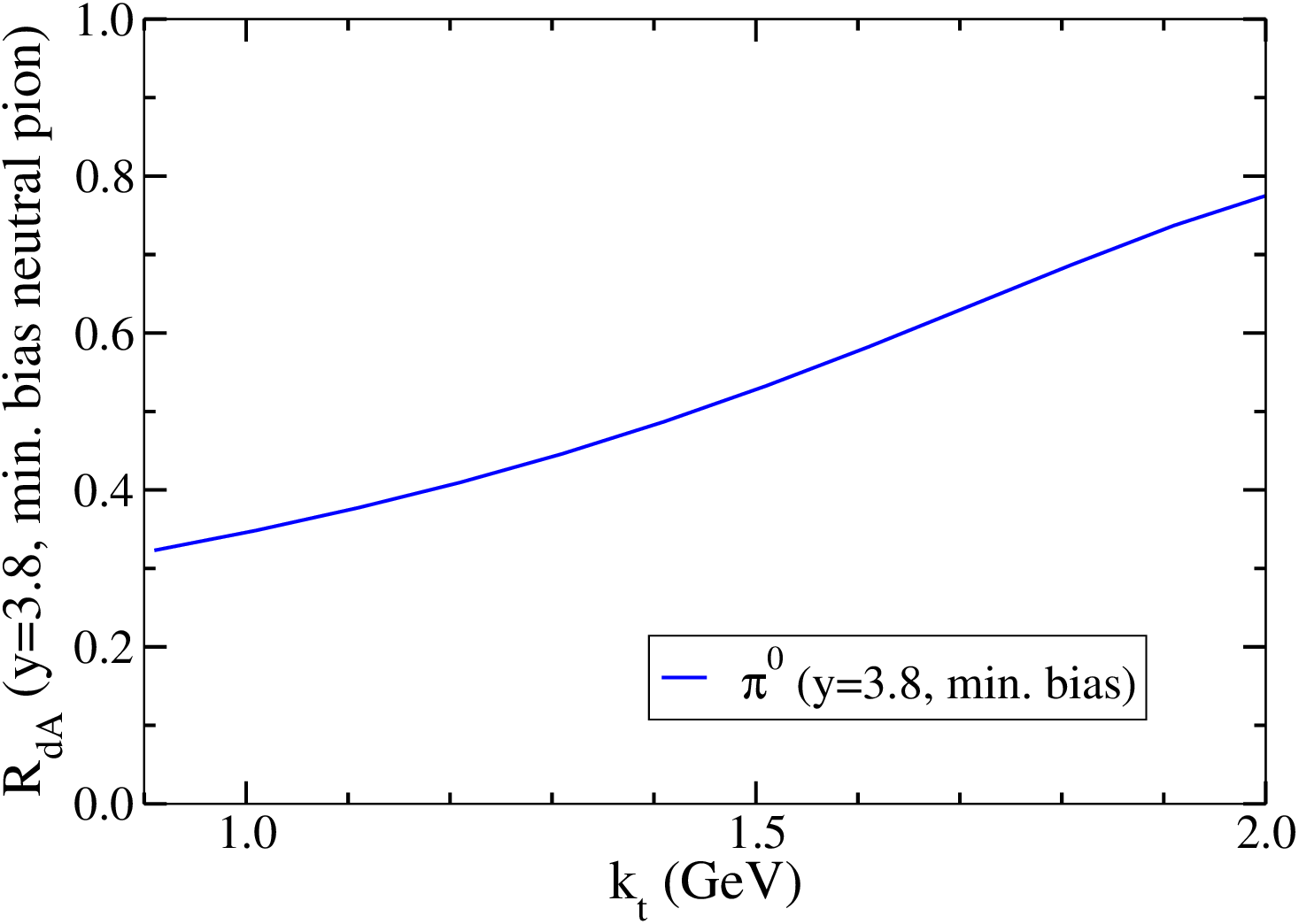}}
\caption{The nuclear modification factor for neutral pions at $y=3.8$ 
in minimum bias deuteron-nucleus collisions at RHIC, $\sqrt{s}=200$ GeV.}
\label{fig:R_dA_38}
\end{figure}

Finally, since the STAR collaboration has measured neutral pions at rapidity
$y=3.8$ in deuteron gold collisions, we show our predictions for neutral
pion nuclear modification factor $R_{dA}$ at $y=3.8$ in 
Fig. (\ref{fig:R_dA_38}). A slightly stronger suppression of 
$R_{dA}$ is seen at lowest $k_t$ as compared to charged hadrons at $y=3.2$.

\section{Discussion}

In deuteron gold collisions in the forward rapidity and low $k_t$ region 
at RHIC considered in this work, the valence quarks are the most abundant 
parton species in the deuteron. They scatter on the target nucleus, 
which has its wave function fully developed (evolved in $x_g$ as much as
allowed by the kinematics) and is characterized by the nucleus saturation 
scale $Q_s^A(x_g)$. The valence quark gets a transverse momentum kick of 
order $Q_s^A$ and is then ``produced''. 

To get a rough idea of the scales involved and estimate where our approach
should break down, we note that hadrons carry $z$ fraction of the parent 
parton energy and that $<z> \sim 0.7-0.8$ in this rapidity (for hadron 
production in $pp$ collisions). The saturation scale of the nucleus is around 
$Q_s^A \sim 1.5 - 1.8$ GeV (minimum bias) at rapidity of $y=3.2$.
The geometric scaling region \cite{ks,iimc} extends to a little higher 
momentum $Q_{es} \equiv Q^2_s/Q_{s0}$ and can be as high as $2.5$ GeV. 
This means that our formalism should describe hadron
production in {\it minimum bias} deuteron gold collisions in the forward 
rapidity region up to $k_t \sim z \, Q_{es} \sim 2$ GeV. We emphasize
that these estimates are for {minimum bias} events only and the Color
Glass Condensate formalism is
expected to be valid at higher $k_t$ for more central collisions.

As one goes to higher $k_t$, gluon radiation becomes important \cite{ykam} 
and will eventually dominate the hadron production cross section. This has 
not been included here and would presumably improve the high $k_t$ 
behavior of the spectra. Another caveat of our approach is the use of 
the dipole parameterization advocated in \cite{iim}. This parameterization 
does not have the Cronin effect \cite{gjm,jmnr,bkw}. This may be partly 
responsible for the deviation of calculated $R_{dA}$ from the data.
Also, this parameterization does not have the right high $k_t$ behavior
since the double log limit is not built into it. This is the reason for 
the decrease of the $R_{dA}$ at higher $k_t$ which is not expected from pQCD.  
However, none of the other simple parameterizations of the dipole 
model \cite{gbw} which have been used to fit the HERA data have the 
right (BFKL) anomalous dimension. Therefore, we use this parameterization
because it has the right anomalous dimension and we are staying in a limited
kinematic region and since it is experimentally known that the Cronin effect 
goes away in the forward rapidity region. 

In order to extend this formalism to mid rapidity, one needs to include
gluon production as considered in \cite{ykam}. The contribution of gluons
to hadron production in mid rapidity is more important than the
contribution of valence quarks since at mid rapidity, one probes small
values of $x$ in the deuteron wave function where there are a lot more
gluons than quarks. Nevertheless, a practical problem with inclusion of 
gluons is that there is no available parameterization of the gluon-gluon 
dipole cross section which has been tested in other processes unlike
the quark anti-quark dipole cross section which is probed in DIS processes. 

An important observable which has not been considered here is the
centrality dependence of the hadron suppression factor in the forward
rapidity region, recently shown by the BRAHMS collaboration in the
Quark Matter $2004$ \cite{brahms}. It will be very interesting to see whether
our formalism can also describe the centrality dependence of the 
data. This would require the application of the Monte Carlo Glauber 
method to our formalism since this is the method used by experimentalists to
extract centrality dependence of the data. This is beyond the scope of 
this work and will be pursued later.

The agreement of our calculations with the forward rapidity data 
becomes even more significant due to the fact we have not used any
free parameters (for $R_{dA}$ and the slope of the spectra) in this 
calculation. All the necessary ingredients,
the dipole cross section (for a proton) and the value of the nucleus 
saturation scale (for minimum bias events) have already been known
for a while and used by various authors \cite{kln,iim}. The fact that 
our simple and {\it parameter free} calculation based on the Color 
Glass Condensate formalism can reproduce the experimental data at 
low $k_t$ is a strong indication that the physics of forward rapidity 
region at RHIC is that of high gluon density QCD and the Color Glass 
Condensate.

\leftline{\bf Acknowledgments} 

We would like to thank F. Gelis, Y. Kovchegov and R. Venugopalan for useful 
discussions. We also thank F. Gelis for providing us with his Fortran code
for Fourier transforming the dipole cross section. This work is supported by 
DOE under grant number DOE/ER/41132.

\leftline{\bf References}

\renewenvironment{thebibliography}[1]
        {\begin{list}{[$\,$\arabic{enumi}$\,$]}  
        {\usecounter{enumi}\setlength{\parsep}{0pt}
         \setlength{\itemsep}{0pt}  \renewcommand{\baselinestretch}{1.2}
         \settowidth
        {\labelwidth}{#1 ~ ~}\sloppy}}{\end{list}}

\end{document}